\documentclass[twocolumn,aps,prl,english,twocolumn,epsfig,groupedaddress]{revtex4}
\usepackage{graphicx}
\usepackage[latin1]{inputenc}
\usepackage{xcolor}
\begin{document}
\title{\bf Unified picture for diluted magnetic semiconductors }

\author{Richard ~Bouzerar$^{1}$\footnote[1]{email: richard.bouzerar@grenoble.cnrs.fr} and Georges ~Bouzerar$^{1,2}$\footnote[6]{email: georges.bouzerar@grenoble.cnrs.fr}
}
\affiliation{
$1.$ Institut N\'eel, CNRS, d\'epartement MCBT, 25 avenue des Martyrs, B.P. 166, 38042 Grenoble Cedex 09, France \\
$2.$ Jacobs University Bremen, School of Engineering and Science, Campus Ring 1, D-28759 Bremen, Germany \\
}
\date{\today}

\begin{abstract}
For already a decade the field of diluted magnetic semiconductors (DMS) is one of the hottest. In spite of the great success of material specific Density Functional Theory (DFT) to provide accurately critical Curie temperatures ($T_{C}$) in various III-V based materials, the ultimate search for a unifying model/theory was still an open issue. Many crucial questions were still without answer, as for example: Why, after one decade, does GaMnAs still exhibit the highest $T_{C}$? Is there any intrinsic limitations or any hope to reach room temperature? How to explain in a unique theory the proximity of GaMnAs to the metal-insulator transition, and the change from RKKY couplings in II-VI materials to the double exchange regime in GaMnN?
The aim of the present work is to provide this missing theory. We will show that the key parameter is the position of the Mn level acceptor and that GaMnAs has the highest $T_{C}$ among III-V DMS. Our theory (i) provides an overall understanding, (ii) is quantitatively consistent with existing DFT based studies, (iii) able to explain both transport and magnetic properties in a broad variety of DMS and (iv) reproduces the $T_{C}$ obtained from first principle studies for many materials including both GaMnN and GaMnAs. The model also reproduces accurately recent experimental data of the optical conductivity of GaMnAs and predicts those of other materials.
\end{abstract}
\parbox{14cm}{\rm}
\medskip

\maketitle

\section{}
In recent years the so called diluted magnetic semiconductors (DMS) have attracted much interest
owing to their potential for spintronic devices \cite{Dietl-science00,Revue-Jungwirth06}. As far as the symbolic barrier of room temperature can be crossed,
their interest is to combine both the traditional electronic functionality and the spin degree of freedom. DMS are obtained by doping a semiconductor host with a little amount of transition metal (TM) ions. Ferromagnetism is mediated by carriers antiferromagnetically coupled to localized spins. Transport and magnetic in DMS properties are very sensitive to dilution effects and to the presence of intrinsic defects. Available theoretical studies can be divided into two different families (i) first principle (DFT) and (ii) model. Note that there are still many dissensions among these numerous studies. First principle based studies were successful to reproduce accurately Curie temperatures for both as grown and annealed samples \cite{GB-APL04,bergqvist04,gbouzerar05a,gbouzerar05b}. Although powerful, first principle methods are essentially fully material-dependent. Thus it is difficult to draw very general conclusions and identify the most relevant physical parameters that control both magnetic and transport properties. Besides DFT studies, the most used model is based on a six or eight realistic bands Kohn-Luttinger (KL) Hamiltonian \cite{Dietl-science00,Revue-Jungwirth06}, including a pd-exchange interaction between localized spins and itinerant holes. However, (i) the pd-coupling is treated perturbatively and (ii) the dilution effects are neglected: this describes the valence band scenario. These studies are inconsistent with first principle calculations which, for example, show the existence of a unambiguous preformed impurity band (IB) in GaMnAs even at relatively high Mn concentration (4-7\%). In addition the VB picture for GaMnAs is unable to explain recent optical conductivity measurements in both as grown and annealed samples \cite{Singley02,Singley03,Burch06}. Finally, those approaches are restricted to metallic systems and can not explain the variations of the Curie temperatures with both hole and magnetic impurity concentrations in III-V compounds. The aim of the manuscript is to propose a model which is able to provide a overall understanding of the magnetic and transport properties both qualitatively and quantitatively in the whole family of III-V materials. The model Hamiltonian (V-J model) that describes the carriers (holes or electrons) interacting with the localized impurity spins and it reads,
\begin{equation}
H=-\sum_{ij, \sigma} t_{ij}c_{i\sigma}^{\dagger}c_{j\sigma}+
\sum_{i}J_{i} {\bf S}_{i}\cdot {\bf s}_{i}+
\sum_{i \sigma}V_{i} c_{i\sigma}^{\dagger}c_{i\sigma}
\label{Hamiltonian}
\end{equation}
The hopping term t$_{ij}$$=$t for nearest neighbors only. $c_{i\sigma}^{\dagger}$ (resp. $c_{i\sigma}$) is the creation (annihilation) operator of a hole of spin $\sigma$ at site i. J$_{i}$ is the local coupling between localized impurity spin {\bf S}$_{i}$ (S$=$5/2 for Mn$^{2+}$) and a spin carrier ${\bf s}_{i}$(p band). Whereas the on-site potential V$_i$ describes the effect of the substitutional disorder. J$_i$$=$p$_{i}$J$_{pd}$ and V$_i$ $=$p$_{i}$V where p$_{i}$$=$1 for impurity sites, otherwise p$_{i}$$=$0. The magnetic impurities are randomly distributed on a simple cubic lattice and $x$ denotes the impurity concentration.
Note that this minimal one band model was already shown to be a good candidate to describe qualitatively the magnetic properties of various III-V DMSs \cite{bouzerar-VJ07,bouzerar-VJ2010}.

\begin{figure}[tbp]
\includegraphics[width=9cm,angle=0]{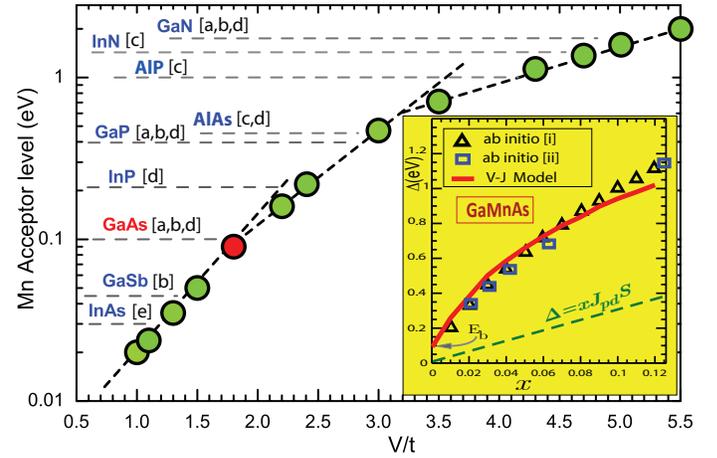}
\caption{(Color online) Calculated Mn acceptor level $E_{b}$ (eV) as a function of V/t in III-V hosts (assuming t$\simeq 0.7~eV$ and J$_{pd}$S$=$4.3~t). The realistic E$_{b}$ [a-e] are extracted from  \cite{Graf02,Zunger04,Masek07,Burch08,InAs}. (Inset) Spin splitting $\Delta$ (eV) as a function of the Mn concentration $\emph{x}$ for Ga$_{1-x}$Mn$_{x}$As. Symbols correspond to LSDA calculations ([i]\cite{Wierzbowska04}, [ii]\cite{Kudrnovsky(b)}) and the continuous line to the V-J model. The dashed line is the MF-VCA expression.
}
\label{splitting}
\end{figure}

\begin{figure}[tbp]
\includegraphics[width=7cm,angle=0]{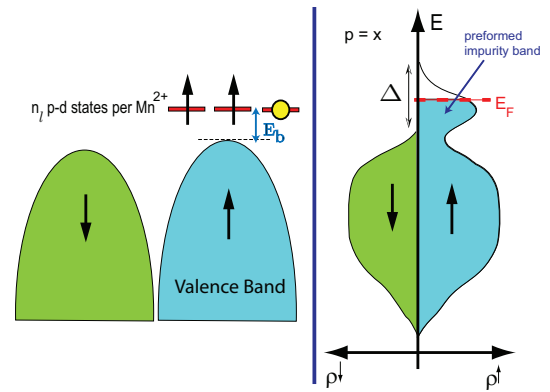}
\caption{(Color online)This cartoon only shows the $n_{l}=3$ pd-states near the top of the valence band for well annealed GaMnAs sample (see left panel). In GaMnAs, each Mn$^{2+}$ brings a localized spin (S$=$5/2) and one hole, thus at finite concentration $x$ the hole density is $p=x$. To be consistent with the 1/3 filled impurity band, within our one band model the calculations are performed for the hole density $\overline{p}=x/3$
}
\label{cartoon}
\end{figure}

We now discuss how we fix this set of 3 parameters. In III-V hosts the bandwidth is rather close to that of GaAs (W$\sim$7-8 eV), therefore, for simplicity, we assume  the same value t$\approx$ 0.7 ~eV for all considered (III,Mn)-V compounds. Note that a variation of about $\pm 10\% $ of t will not affect considerably our results. In both (II,Mn)VI or (III,Mn)-V DMS J$_{pd}$ is about 1 eV \cite{Jpd1,Jpd2}, thus in the following we set J$_{pd}$$=$1.2 eV (accepted value for GaMnAs).
Thus, a single parameter is left (V). This parameter V is set to reproduce for each compound the bound hybridized pd states energy (E$_{b}$) with respect to the top of the VB.
Note that this bound-state is three-fold degenerate, this is illustrated in the left panel of Fig.\ref{cartoon}. At finite, but small concentration on Mn and because each Mn provides a single hole this will lead to an impurity band 1/3 hole filled (see right panel of Fig.\ref{cartoon})\cite{Sanvito02,Bergqvist03,Sandratskii04} .
In the case of a single Mn in GaAs the bound states lies at E$_{b}$ $\approx$ 110 ~meV \cite{ChapLin,Yakuchen} above the VB.
In Fig.\ref{splitting}, the calculated E$_{b}$ is plotted as a function of V/t. In this figure the realistic values of Mn acceptor level in various hosts (GaAs,InAs,InN..) are shown.
These values were obtained either experimentally or within first principle calculations \cite{Graf02,Zunger04,Masek07,Burch08,InAs}.
For GaMnAs we have found V$=$1.8 t and 5 t for GaMnN. The inset of Fig.\ref{splitting} shows the calculated spin splitting $\Delta(x) = E_{max}^{\uparrow} -E_{max}^{\downarrow}$ for GaMnAs as a function of $x$, $E_{max}^{\sigma}$ is the largest eigenvalue in the $\sigma$ sector (see right panel of Fig.\ref{cartoon} for an illustration). We see that our results are in excellent agreement with those obtained from ab initio LSDA calculations \cite{Wierzbowska04,Kudrnovsky(b)} (open symbols). The dashed line shows the Mean Field Virtual Crystal Approximation (MF-VCA) value
$\Delta$=$x$J$_{pd}$S. In contrast to what is often found in the literature, the MF-VCA expression can not be used to extract J$_{pd}$ in III-V DMS. For that reason
the value that was found in \cite{Wierzbowska04} was much larger (4 times) than the value of 1.2 eV. However, $\Delta$=$x$J$_{pd}$S is a good approximation in II-VI materials as ZnMnTe or CdMnTe (small $|V|$ values).
In our one band model each Mn$_{Ga}$ brings a single state. Thus, to be consistent with the realistic compound in which each Mn provides a single hole and $n_{l}=3$ p-d states, our model calculations for well annealed samples will performed at the hole density $\overline{p}=x/3$. We will get back to this important point in the following.

\par
Let us now first discuss the procedure to calculate the magnetic properties in any compound characterized by its set of 3 parameters. For each configuration of disorder (distribution of magnetic impurities in the host) we diagonalize exactly the Hamiltonian (\ref{Hamiltonian}) in both spin sectors assuming fully polarized d-spins (for details see ref.\cite{bouzerar-VJ07}). Typically we have used simple cubic systems size from $16^{3}$ to $24^{3}$ sites and the average are done with a few hundreds of disorder configurations.
The diagonalization for each disorder configuration (denoted c) provides the full spectrum, eigenvalues and eigenvectors denoted \{$E_{\sigma,\alpha}^{c}$, $|\Psi\rangle_{\sigma,\alpha}^{c} $\} ($\alpha$ denotes the eigenstate index) needed to calculate magnetic couplings and transport properties.
The magnetic couplings between two localized spins, respectively at site i and j, is given by the generalized susceptibility
\begin{equation}
 \overline{{J}}_{{i},{j}}(x,\overline{p})= -{\frac {1} {4\pi S^2}} \Im \int_{-\infty}^{E_F} Tr(\Sigma_{{i}}G_{{i},{j}}^{\uparrow}(\omega) \Sigma_{{j}} G_{{j},{i}}^{\downarrow}(\omega)) d\omega\nonumber
\label{Echange}
\end{equation}
where the Green's function are G$_{i,j}^{\sigma}(\omega)=\langle i\sigma|\frac{1}{\omega -\hat{H} +i\epsilon} | j\sigma \rangle$. Within our model the local exchange splitting reduces to $\Sigma_{i}=$J$_{pd}$S.
As mentioned before, since our model provides a single state per impurity (instead of 3 p-d states in the realistic approach) then the magnetic exchange that have to be used to calculate the Curie temperature are defined as ${J}_{{i},{j}}(x,p)=n_{l}\overline{{J}}_{{i},{j}}(x,\overline{p}=p/n_{l})$. We note that the same argument applies for the transport properties. In other words, the optical conductivity will be defined as ${\sigma}(\omega,p)=n_{l}\overline{{\sigma}}(\omega,p/n_{l})$ where $\overline{{\sigma}}(\omega,p/n_{l})$ is calculated within our one band model, the results will be discussed in the last section. Note that the overall shape of the couplings agree well with those obtained from first principle studies. For example, for Ga$_{1-x}$Mn$_{x}$As they are rather short range and essentially ferromagnetic for well annealed compounds \cite{bouzerar-VJ07,Kudrnovsky04}. To calculate the Curie temperature, the effective dilute Heisenberg Hamiltonian $H_{Heis}=-\sum_{i,j} J_{ij}(x,p) \textbf{S}_{i}\cdot\textbf{S}_{j}$ (sums runs over sites occupied by Mn$^{2+}$) is solved within the self consistent local RPA theory (SCLRPA)\cite{gbouzerar05a}. Note also that we could also calculate other magnetic properties such as magnetization as a function of T, magnons spectrum, spin stiffness ... but this is not the scope of the present work. Note also that the accuracy and reliability of SC-LRPA was already demonstrated several times.

\begin{figure}[tbp]
\includegraphics[width=8cm,angle=00]{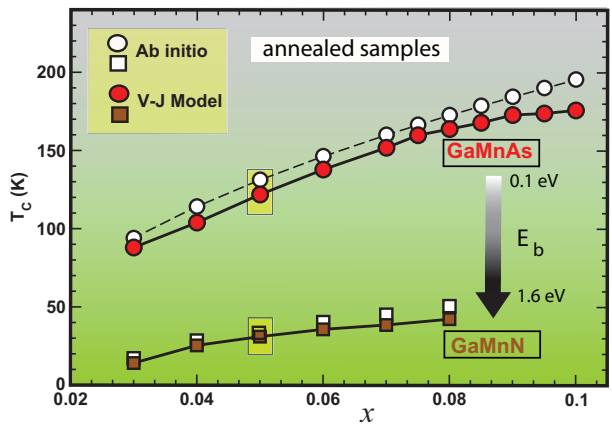}
\caption{(Color online) Curie temperature (in K) for Ga$_{1-x}$Mn$_{x}$As and
Ga$_{1-x}$Mn$_{x}$N as a function of Mn concentration $x$ within both ab-initio \cite{gbouzerar05a} (open symbols) and model calculations (filled symbols).
}
\label{Tc-GaMnAs-GaMnN}
\end{figure}

In Fig.\ref{Tc-GaMnAs-GaMnN} we have plotted the Curie temperature (in Kelvin) for both
Ga$_{1-x}$Mn$_{x}$As and Ga$_{1-x}$Mn$_{x}$N as a function of Mn concentration assuming well annealed samples (p=$x$). Surprisingly, the agreement between model and first principle based calculations \cite{gbouzerar05a,bergqvist04,Sato04} is not only qualitative but also quantitative in the overall range of Mn concentration. It is important to stress again that no fitting parameter is used. Note that the ab initio results were themselves in very good agreement with the experimental data \cite{Edmonds,Matsukura98,Chiba03}. These results already clearly demonstrate that our model captures accurately the physics in these materials, this will be confirmed in the following. Note that the nature of the couplings are very different in these two compounds. Indeed, in GaMnN the couplings are of double exchange type whilst more extended in the case of GaMnAs. Note also that a variation of about $\pm 10\% $ of the hopping t does not affect considerably the quantitative results (box symbol for $x=0.05$).
\begin{figure}[tbp]
\includegraphics[width=9cm,angle=0]{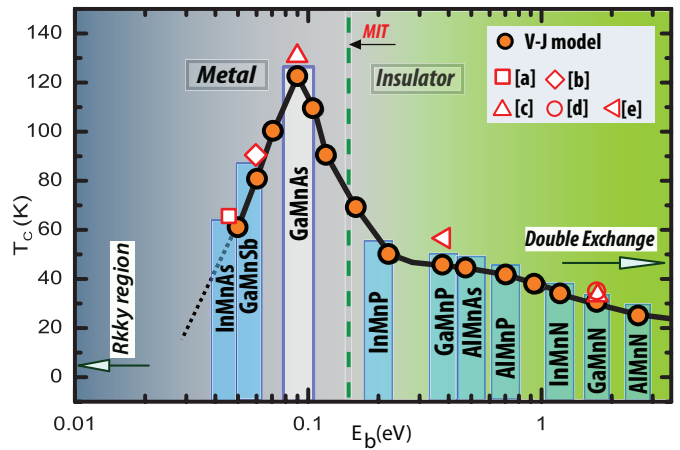}
\caption{(Color online) Curie temperature (in K) as a function of the bound state $E_{b}$ (eV) and for a Mn concentration $x=5\%$ in well annealed samples. The filled symbols correspond to our model, the open symbols to $T_{C}$'s computed with ab initio exchange integrals ([a-d] are respectively from Refs.\cite{bergqvist08},\cite{gbouzerar-unpub},\cite{gbouzerar05a},\cite{Sato04}). }
\label{Tc(V)}
\end{figure}

\par
In Fig.\ref{Tc(V)} we now show the model calculations of the Curie temperatures (in K) as a function of E$_{b}$ (eV) for various (III,Mn)V compounds. We present only the case of $x=5\%$ of substitutional Mn and a hole density p$=$x. In this figure, available T$_{C}$'s obtained from ab-initio exchange integrals \cite{gbouzerar05a,Sato04,bergqvist08,gbouzerar-unpub} are also plotted. Again we obtain a very good agreement between the V-J model calculations and those obtained starting from first principle approaches.
The model calculations show that the T$_{C}$(E$_{b}$) curve exhibits a clear and pronounced resonant peak structure and predicts that Ga$_{1-x}$Mn$_{x}$As has the highest critical temperature among III-V Mn doped materials. By mean of transport calculations, the mobility edge is found close to the value V$=$2.1~t (E$_{b}$$\approx$ 0.15~eV).
Interestingly, the narrow resonant peak (30 $\le$ E$_{b}$ $\le$ 200 ~meV) appears to be located near the metal-insulator phase transition. Thus, this already indicates why experimentally both metallic and insulating compounds were obtained and why after annealing as-grown samples the materials exhibit an insulator-metal transition. For small E$_{b}$ (or V), the couplings are RKKY like and thus the Curie temperatures are very small \cite{bouzerar-RKKY06,bouzerar-VJ07}. This region corresponds to II-VI materials as ZnMnTe or CdMnTe. As E$_{b}$ increases the critical Curie temperature gets larger resulting from the suppression of the RKKY oscillations and the couplings become shorter range (resonant effects due to the preformed impurity band). By increasing further V (thus E$_{b}$ ) the range of the couplings becomes shorter and shorter and the relevant couplings that controls the critical temperature get smaller, thus T$_C$ reduces too. For very large $E_{b}$ the couplings become double exchange like leading to much smaller T$_C$. In this region the physics of percolation becomes more crucial (included in our theory). This is the case of GaMnN or AlMnN in which the nearest neighbor coupling dominates strongly. Within our theory Ga$_{0.95}$Mn$_{0.05}$P is found to be insulator in agreement with experimental observations \cite{Burch08}. Note also that our predictions for Ga$_{0.95}$Mn$_{0.05}$P and In$_{0.95}$Mn$_{0.05}$As (respectively $\approx$ 50 K and 60 K) are close to the experimental maximum measured value ($T_{C}^{exp}\approx 60 ~K$ for annealed samples with $x\approx 6 \%$) \cite{Wang06,Ohno03,Scarpulla05,Burch08}.

\par
\begin{figure}[tbp]
\includegraphics[width=8cm,angle=00]{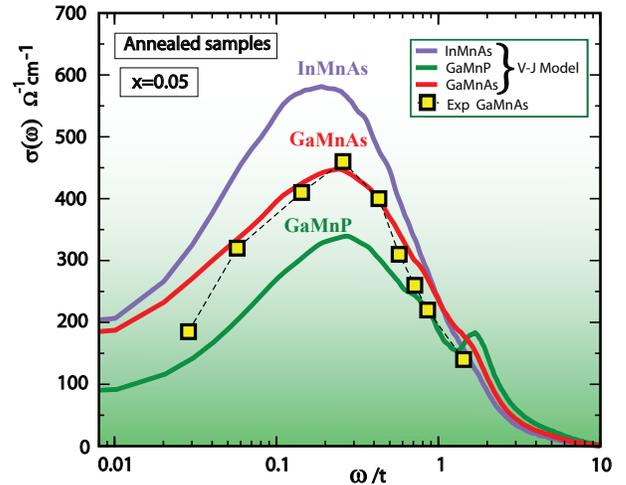}
\caption{(Color online)  Model calculation of the conductivity (in unit $\Omega^{-1}\cdot cm^{-1}$) as a function of $\omega/t$ for well annealed ($p=x$) In$_{0.95}$Mn$_{0.05}$As, Ga$_{0.95}$Mn$_{0.05}$As and Ga$_{0.95}$Mn$_{0.05}$P. The symbols correspond to experimental measurements in Ga$_{0.95}$Mn$_{0.05}$As \cite{Burch06}.}
\label{Fig5}
\end{figure}

 In this section we now discuss whether our theory is also successful to explain transport measurements. We have performed the optical conductivity as a function of the frequency $\omega$ in well annealed In$_{1-x}$Mn$_{x}$As, Ga$_{1-x}$Mn$_{x}$As and Ga$_{1-x}$Mn$_{x}$P for $x=5\%$.
 As discussed above, the optical conductivity is defined by ${\sigma}(\omega,p)=n_{l}\sum_{\sigma}{\overline{\sigma}}_{\sigma}(\omega,p/n_{l})$. The regular part of the one band optical conductivity is given within the Kubo formalism by,
 \begin{equation}
 {\overline{\sigma}}_{\sigma}(\omega,p/n_{l})=\frac{1}{N} \frac{{\pi e^2}}{\hbar a}\sum_{\alpha\neq\beta}(n^{\sigma}_{\alpha}-n^{\sigma}_{\beta})A_{\alpha,\beta}^{\sigma}\delta(\hbar\omega-E^{\sigma}_{\alpha}+E^{\sigma}_{\beta})
\label{kubo}
\end{equation}
$N=L^{3}$ is the total number of sites, $a$ is the simple cubic lattice parameter. The lattice parameter of GaAs fcc lattice is $a_{0}\simeq 5.65 ~10^{-10}$ m (4 atoms per unit cell), to keep the same unit cell volume a is state to $a=\frac{a_{0}}{4^{1/3}}$. The matrix element $A_{\alpha,\beta}^{\sigma}=\frac{|\langle \Psi^{\sigma}_{\alpha}|\hat{j}_{x}|\Psi^{\sigma}_{\alpha} \rangle |^{2}}{E^{\sigma}_{\alpha}-E^{\sigma}_{\beta}}$, where $\hat{j}_{x}$ is the current operator in the $x$ direction and $n^{\sigma}_{\alpha}$ is the occupation number of the state $|\Psi^{\sigma}_{\alpha}\rangle$. In Fig.\ref{Fig5} we compare the results of our model calculations with available experimental data for Ga$_{0.95}$Mn$_{0.05}$As \cite{Burch06}. We find an excellent quantitative agreement on the whole frequency range. In particular the peak location at $0.2~eV$ is accurately reproduced. Note that we have recently performed a detailed analysis of the effects of compensating defects on the transport properties of GaMnAs \cite{bouzerarGB-VJ2010} in presence of compensating defects and it leads to excellent quantitative agreement with recent experimental data \cite{Burch06}. In particular our theory is in agreement with the observed red-shift of the broad peak while increasing the hole density from compensating to annealed GaMnAs samples. This feature is in agreement with the fact that Ga$_{0.95}$Mn$_{0.05}$As exhibits a preformed impurity band. In the same figure, we also predict some trends for Ga$_{0.95}$Mn$_{0.05}$P and In$_{0.95}$Mn$_{0.05}$As. At low frequency the conductivity of InMnAs is seen to be higher than the two other compounds which is correlated with the fact that the Fermi level in InMnAs is deeper in the valence band (more extended states), let us recall that the Mn level acceptor lies at E$_{b}\approx 300 ~meV$. For this material the optical conductivity peak is located at $0.11~eV$. For Ga$_{0.95}$Mn$_{0.05}$P the shape of ${\sigma}(\omega)$ exhibits two clearly distinct peaks. The principal one is located at $\omega\simeq0.2~eV$ and the narrower secondary peak lies at $\omega\simeq1.2~eV$. For this Mn density it is interesting to notice that GaMnP is at the edge where the impurity band separates from the valence band. It's secondary peak originates from electronic transitions between the separated impurity band and the top of the VB. As we move to larger $E_{b}$ we expect a shift of this secondary peak to highest frequency (case of GaMnN for example). For these last compounds it should be interesting to have transport measurements in order to confirm or infirm these predictions.

\par
To conclude, we have  drawn up a model for diluted magnetic semiconductors based on the minimal V-J Hamiltonian  treated in a non perturbative way and without the use of effective medium. The parameters of the model are fixed in agreement with first principle band calculation. For example, concerning well annealed GaMnAs (and without inhomogeneities), that process allowed us to reproduce quantitatively the Curie temperatures obtained from first principle study which were found to be in excellent agreement with the experimental values. While varying the local potential V (or equivalently the bound state $E_{b}$), the resonant structure of the critical temperature shows that GaMnAs can be considered as optimal among the (III,Mn)-V family.  Moreover our theoretical model is in excellent agreement with recent measurements of the optical conductivity for GaMnAs. For the band gap DMS GaMnN, the critical temperatures obtained from our model are also in good agreement with first principle calculations. For the first time a theoretical model allow to describe qualitatively and quantitatively both transport and magnetic properties of a broad range of diluted magnetic semiconductors and on both sides of the metal-insulator phase diagram. This unifying picture bridges the gap between model approach and first principle calculations. Successes of this model, show that it represents a tool of choice  to understand experimental data. In addition it allows interesting predictions for specific materials.

\section*{Acknowledgments}
\phantomsection

We would like to thank E.Kats for providing us with ILL computer facilities and M. Johnson for the access to the CS group's clusters.

\addcontentsline{toc}{section}{Acknowledgments}


\end{document}